\documentclass[aps,twocolumn,pra,groupedaddress,amsmath,amssymb,10pt,floatfix]{revtex4-2}
\usepackage{dcolumn}
\usepackage{bm}
\usepackage{amsmath}
\usepackage{txfonts}
\usepackage[T1]{fontenc}
\usepackage{xspace}
\usepackage{ulem}
\usepackage{comment}
\usepackage{braket}
\usepackage{bm}
\usepackage{physics}
\usepackage{comment}
\ifx\pdfoutput\undefined
\usepackage[dvipdfmx]{graphicx}
\usepackage[dvipdfmx]{hyperref}
\usepackage[dvipdfmx]{color}
\else
\usepackage{graphicx}
\usepackage{hyperref}
\usepackage{color}
\fi
\newcommand{\add}[1]{#1}
\begin{document}
\let\emph\textit
\title{Itinerant ferromagnetism in an SU(3) Fermi-Hubbard model at finite temperatures: A dynamical mean-field theory study}

\author{Juntaro Fujii}
\email{fujii.j.47dc@m.isct.ac.jp}
\author{Kazuki Yamamoto}
\author{Akihisa Koga}
\affiliation{
Department of Physics, Institute of Science Tokyo, Meguro, Tokyo 152-8551, Japan
}
\altaffiliation{Former Tokyo Institute of Technology}

\date{\today}
\begin{abstract}
We investigate an SU(3) Fermi-Hubbard model on a hypercubic lattice at finite temperatures, combining dynamical mean-field theory with continuous-time quantum Monte Carlo simulations. Taking strong correlations into account carefully, we find a ferromagnetically ordered state, in which one of the three components becomes dominant, when holes are doped away from one-third filling. Furthermore, we demonstrate that this ferromagnetically ordered phase undergoes a first-order transition to a paramagnetic state. We clarify the stability of the ferromagnetically ordered state against interaction strength, hole doping, and temperatures. The relevance of generalized Nagaoka ferromagnetism is also addressed, by comparing the results on the Bethe lattice.
\end{abstract}
\maketitle

\section{Introduction}\label{sec:i}
Experimental advancement with ultracold atoms in optical lattices has opened a new avenue for studying exotic quantum many-body phenomena \cite{takahashi_2020}. These systems offer rich possibilities for quantum simulations of strongly correlated systems

\add{~\cite{greif2013short,hart2015observation,boll2016spin,parsons2016site,cheuk2016observation,he2019recent,lebrat2024observation}}, 
even when the parameter tuning is difficult in solids. For example, the interatomic interactions are controlled by using the Feshbach resonance~\cite{Chin10} and the dimensionality~\cite{grimm2000optical} can also be controlled by tuning the trap lasers. 

\add{Recently, by utilizing the hyperfine states of nuclear spins in ultracold fermions such as $^6$Li~\cite{ottenstein_collisional_2008,Huckans2009three}, $^{87}$Sr~ \cite{desalvo_degenerate_2010,zhang2014spectroscopic}, $^{173}$Yb~\cite{Fukuhara2007degenerate,taie2010realization,pasqualetti2024EOS}, and $^{40}$K~\cite{krauser2012coherent, krauser2014giant}, }
multicomponent Fermi gases including the SU($N$) Fermi-Hubbard model with $N>2$~\cite{hubbard} have been realized. Experimentally, a variety of phenomena have been achieved in SU($N$) quantum systems~\cite{taie2010realization,taie_su6_2012,scazza_observation_2014,zhang2014spectroscopic, tusi_flavour_selective_2022, Mong25}, such as the antiferromagnetic spin correlations of the SU($N$) Fermi gas~\cite{ozawa_antiferromagnetic_2018,taie_observation_2022} and the crossover from metals to Mott insulators with exotic compressibility~\cite{hofrichter_direct_2016}. These studies highlight the novel many-body physics emerging from multicomponent spins, facilitating theoretical investigations of rich SU($N$) physics~\cite{congjun2003exact,carsten2004ultracold,Michael2009Mott,Rapp2011ground_state,cazalilla_ultracold_2014,Sotnikov2014magntic, yanatori_finite_2016, capponi_phases_2016,Yoshida2021rigorous,yamamoto_universal_2023,Feng_2023,Nakagawa2024exact,ibarra_many_body_su_N_2024}.

On another front, itinerant ferromagnetism is the prototypical strongly correlated phenomenon in condensed matter physics and has attracted great attention~\cite{nagaoka_1965, nagaoka_1966,thouless_exchange_1965, lieb_theory_1962,tasaki_extension_1989,lieb_two_1989,tasaki_ferromagnetism_1992,mielke_ferromagnetism_1993,riera_ferromagnetism_1989,obermeier_ferromagnetism_1997,white_density_2001,zitzler_magnetism_2002,park_dynamical_2008,peters_magnetic_2009,liu_phases_2012,kamogawa_ferromagnetic_2019,Koga_pro2021,yun_ferromagnetic_2023,samajdar_polaronic_2024,schlomer2024kinetic}. An earlier study by Nagaoka~\cite{nagaoka_1965, nagaoka_1966} rigorously showed that, in a lattice with closed loops, the ground state of the infinite-$U$ SU(2) Fermi-Hubbard model with a single hole doping is a fully polarized ferromagnetic state, which arises from the itinerant motion of the hole. Furthermore, the effect of interactions and hole doping on ferromagnetism was numerically 
investigated
\add{~\cite{obermeier_ferromagnetism_1997,zitzler_magnetism_2002,park_dynamical_2008,peters_magnetic_2009,kamogawa_ferromagnetic_2019,Koga_pro2021,schlomer2024kinetic,kim2024itinerant}. }
Itinerant ferromagnetism in the SU$(N)$ Fermi-Hubbard model with $N>2$ has been actively investigated in recent decades~\cite{cazalilla_2009,nie_2017,bobrow_exact_2018,liu_flat_band_2019,tamura_ferromagnetism_2021,singh_divergence_2022,tamura_flat_2023,huang_2023,botzung_numerical_2024}. In the ferromagnetically ordered state in the SU($N$) system, flavor imbalance emerges among the $N$ components. A special case is known as the polarized state where one of the flavors is dominant~\cite{cazalilla_2009,katsura_2013}. It has been rigorously shown that the fully polarized state is realized in the infinite-$U$ limit with a single hole doping at the $1/N$ filling~\cite{katsura_2013,tamura_ferromagnetism_2021,tamura_flat_2023}, which is regarded as the SU($N$) generalization of Nagaoka's theorem. 

\add{Notably, it has been clarified that the ferromagnetic phase transition is of first
order in the SU($N$) systems with $N > 2$, in contrast to the second-order
transition in the SU(2) case. This qualitative difference can be understood
within symmetry arguments based on Landau theory~\cite{cazalilla_2009}.}
Moreover, it has been clarified that the critical interaction strength is enhanced as $N$ increases~\cite{botzung_numerical_2024}. 
\add{
These results demonstrate that the nature of SU($N$) ferromagnetism differs from that of the conventional SU(2) system.
However, describing the ferromagnetically ordered state
realized in the strong-coupling regime
and evaluating its stability against thermal fluctuations
requires accounting for strong dynamical electron correlations
beyond the simple methods such as static mean-field approximation~\cite{cazalilla_2009}
and the exact diagonalization~\cite{botzung_numerical_2024}.
}
Therefore, it is important to quantitatively examine the stability of the ferromagnetically ordered state in the SU($N$) system by means of numerical approaches. 

In this paper, we investigate itinerant ferromagnetism in the doped SU(3) Fermi-Hubbard model on a hypercubic lattice using dynamical mean-field theory (DMFT)~\cite{metzner_correlated_1989,muller_hartmann_correlated_1989,georges_dynamical_1996}, where the continuous-time quantum Monte Carlo (CT-QMC) method~\cite{werner_continuous_2006,gull_continuous_2011} with a nonuniform sampling scheme is used as an impurity solver. We demonstrate that the ferromagnetically ordered state, which lies at low temperatures and under strong correlations, is realized as a result of spontaneous symmetry breaking. We find that the transition between the ferromagnetically ordered and the paramagnetic states is of first order. Furthermore, we demonstrate that the ferromagnetically ordered state is stabilized at lower temperatures and requires stronger correlations compared to the SU(2) case.

The rest of this paper is organized as follows. In Sec.~\ref{sec:mm}, we introduce the SU(3) Fermi-Hubbard model and provide a DMFT framework with the numerical improvement together with the introduction of physical quantities. In Sec.~\ref{sec:r}, by examining magnetization, magnetic susceptibility, and energy, we clarify that the ferromagnetically ordered state is realized at low temperatures. Finally, the summary is given in Sec.~\ref{sec:conclusion}.

\section{Model and methods}\label{sec:mm}

\begin{figure}[htbp]
  \includegraphics[width=0.9\linewidth]{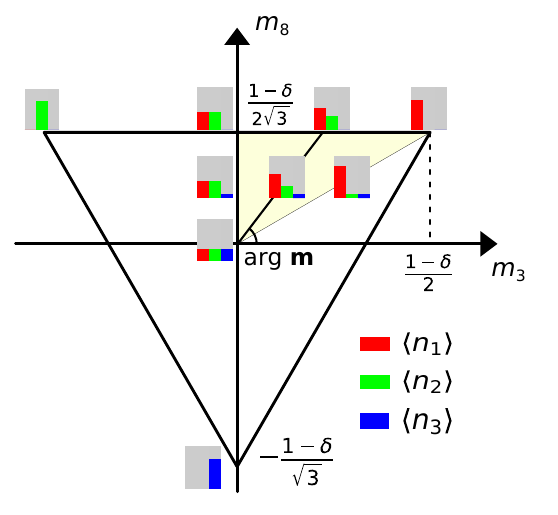}
  \caption{
  Triangular region in the ${\bf m} = (m_3, m_8)$ plane represents possible magnetizations for the ferromagnetically ordered state in the doped system. Red, green, and blue bars represent the particle number of each flavor $\langle n_1 \rangle$, $\langle n_2 \rangle$, and $\langle n_3 \rangle$, respectively.
  }
  \label{A_aa}
\end{figure}

We consider the SU(3) Fermi-Hubbard model on a hypercubic lattice. The Hamiltonian is given by
\begin{equation}
      H = -t \sum_{\langle i,j \rangle, \sigma} c_{i,\sigma}^{\dagger} c_{j,\sigma}  + \frac{U}{2}\sum_{i, \sigma \neq \sigma'} n_{i,\sigma}n_{i,\sigma'}, 
  \label{hamiltonian}
\end{equation}
where $c_{i,\sigma}^{\dagger}$ ($c_{i,\sigma}$) denotes the creation (annihilation) operator of the fermion with flavor $\sigma(=1,2,3)$ at site $i$, $n_{i,\sigma} = c_{i,\sigma}^{\dagger}c_{i,\sigma}$. Here, $t$ is the hopping amplitude between nearest-neighbor sites $\langle i,j \rangle$ and $U$ is the repulsive on-site interaction. This system exhibits SU(3) symmetry as the Hamiltonian commutes with the flavor operators $S_\alpha\;(\alpha=1,2,\ldots, 8)$~\cite{katsura_2013,nie_2017}, defined as 
\begin{align}
  S_\alpha=\frac{1}{2}\sum_{i,\sigma, \sigma'} c_{i,\sigma}^{\dagger} (\lambda_{\alpha})_{\sigma, \sigma'} c_{i, \sigma'},
  \label{eq_x}
\end{align}
where $\lambda_\alpha$ denotes the $\alpha$th Gell-Mann matrix~\cite{gell_mann_symmetries_1962}. The SU(3) Hubbard model has been extensively discussed, and interesting many-body phenomena have been found such as magnetically ordered states
\add{~\cite{sotnikov_2015,yanatori_finite_2016,nie_2017,Chung_2019,ibarra2023metalinsulator,Henning2024sudimensional,zhang2025unit,kleijweg2025zigzag,bohler2025magnetic}}
and Mott transitions~\cite{Assaraf_1999,Mott_2005,Manmana_2011,Blumer2013Mott,Lee_2018,HafezArtifical2018artifical,Feng_2023}.

To study the SU(3) Fermi-Hubbard system, we employ DMFT~\cite{metzner_correlated_1989,muller_hartmann_correlated_1989,georges_dynamical_1996}, which is one of the most powerful approaches for treating strongly correlated systems. In this framework, the original lattice model is mapped to a single impurity model connected to an effective bath, where dynamical correlations are precisely accounted for. As DMFT is exact in infinite dimensions, and even in three dimensions, it is expected to capture essential features of many-body phenomena. Indeed, DMFT has been widely applied to the multicomponent system to explain interesting physics such as Mott  transitions~\cite{Kotliar_1996,Rozenberg_1997,Han_1998,Koga_2002,Ono_2003,koga_2004,koga_2005}, superconductivity~\cite{Capone_2001,Rapp_2007,Inaba_2009,Hoshino2015supercunductivity,koga_2015,ishigaki_2018,Okanami2014Stability,Yue_2021}, and magnetism~\cite{Momoi_1998,Held_1998,yanatori_finite_2016,Koga_2017}.

In our study, we focus on magnetic properties below one-third filling.

The density of doped holes is given by $\delta = 1 - \sum_{\sigma} \langle n_{\sigma} \rangle$ and $n_{\sigma} = \sum_{i} n_{i,\sigma} / L $, where $L$ is the total number of sites and $\langle A\rangle$ denotes the expectation value of the operator $A$. 
\add{
We fix the doping rate by adjusting the chemical potential $\mu$ in the framework of DMFT.
} 

Since a ferromagnetically ordered state is considered with a spatially uniform particle distribution, one can choose $\lambda_3$ and $\lambda_8$ as the principal axes of the SU(3) weight diagram, where $\lambda_3 = \mathrm{diag}(1,-1, 0)$ and $\lambda_8 = \mathrm{diag} (1,1,-2) /\sqrt{3}$.  

In this case, the magnetization is represented in two dimensions as ${\bf m}=(m_3, m_8)$ with $m_{\alpha}=\langle S_{\alpha}\rangle/L$. Figure~\ref{A_aa} shows the flavor configurations in the ${\bf m}$ plane. When $\mathrm{arg}\,{\bf m }= \pi/6\,\, \mathrm{mod}\,(2\pi/3)$, only one flavor exhibits a dominant occupation. Conversely, at $\mathrm{arg}\,{\bf m } = \pi/2\,\, \mathrm{mod}\,(2\pi/3)$, the particle occupations in two of the three flavors are larger than in the remaining one. In the following, we focus on the region $\pi/6\leq \mathrm{arg}\,{\bf m }\leq\pi/2$ without loss of generality. 
\add{As we take the principal axes along the diagonal elements of the matrix, we focus solely on diagonal terms of the Green's function and the self-energy. Then, the Dyson equation for the lattice Green's function is given by}
\begin{equation}
 G^{\sigma}(k,i\omega_n)^{-1} = i\omega_n+\mu-\epsilon_k-\Sigma^{\sigma}(k, i\omega_n),
\end{equation}
where $\epsilon_k$ is the dispersion relation, $\mu$ is the chemical potential, $\omega_n[=(2n+1)\pi T]$ is the Matsubara frequency, and $T$ is the temperature. $G^{\sigma}(k,i\omega_n)$ and $\Sigma^{\sigma}(k, i\omega_n)$ are the lattice Green's functions and the self-energy, respectively. In infinite dimensions, the self-energy is momentum-independent, $\Sigma^{\sigma}_{\mathrm{loc}}(i\omega_n) = \Sigma^{\sigma}(k, i\omega_n)$. The local Green's function is given by 
\begin{align}
  G_{\text{loc}}^{\sigma}(i\omega_n) 
  &= \int \dfrac{\rho(\epsilon)\text{d}\epsilon}{i\omega_n +\mu -\epsilon - \Sigma_{\mathrm{loc}}^{\sigma}(i\omega_n)},
  \label{eq_lattice_l}
\end{align}
with
\begin{align}
  \rho(\epsilon) 
  &= \frac{1}{\sqrt{\pi}\,D}\,\exp\!\left[-\left(\frac{\epsilon}{D}\right)^2\right],
\end{align}
where $D$ is the bandwidth of the noninteracting density of state $\rho$.
In the effective impurity model, the effective bath is also flavor independent. Therefore, it is described by
\begin{equation}
    \mathcal{G}^{\sigma}(i\omega_n)^{-1} = i\omega_n+\mu-\Delta^\sigma(i\omega_n),
    \label{eq_imp}
\end{equation}
\add{where $\Delta^\sigma(i\omega_n)$ is the hybridization function and $\mathcal{G}^{\sigma}(i\omega_n)$ is the Green's function of the effective bath~\cite{georges_dynamical_1996}. It should be noted that the hybridization function is determined by the self-consistency equations. 
}

By solving the effective impurity model, we obtain the Green's function $G_{\text{imp}}^{\sigma}(i\omega_n)$ and
self-energy $\Sigma_{\mathrm{imp}}^{\sigma}(i\omega_n)$. The self-consistent equations are $G^\sigma_{\text{imp}}(i\omega_n)=G_\mathrm{loc}^\sigma(i\omega_n)$ and $\Sigma_{\mathrm{imp}}^{\sigma}(i\omega_n) = \Sigma_{\mathrm{loc}}^{\sigma}(i\omega_n)$. We numerically solve the effective impurity problem, then update the Green's function and self-energy, and iterate this self-consistent procedure until the result converges within numerical accuracy.

In our DMFT calculations, we use the hybridization-expansion CT-QMC method employing the segment algorithm as an impurity solver~\cite{werner_continuous_2006,gull_continuous_2011}. This method efficiently samples Monte Carlo configurations through local updates, such as inserting or removing segments and anti-segments, or shifting segment endpoints. However, when $U\gg t$, two major challenges arise. First, the acceptance probability drops exponentially, making it difficult to compute the Green's function efficiently. To address this, we use an additional scheme that simultaneously updates the configurations of two flavors within a selected time interval, which significantly improves sampling efficiency in the strong-coupling regime~\cite{koga_low_temp_2011}. Second, the Green's function is difficult to represent accurately due to the presence of widely separated energy scales that lead to distinct decay behavior in imaginary time. To overcome this, we employ intermediate-representation basis functions~\cite{shinaoka_compressing_2017,chikano_irbasis_2019} combined with a nonuniform sampling scheme~\cite{li_sparse_2020}, which enables accurate and efficient reconstruction of the Green's function. Some details are provided in Appendix~\ref{non_uniform_sampling}.

\begin{figure}[htb]
\includegraphics[width=0.9\linewidth]{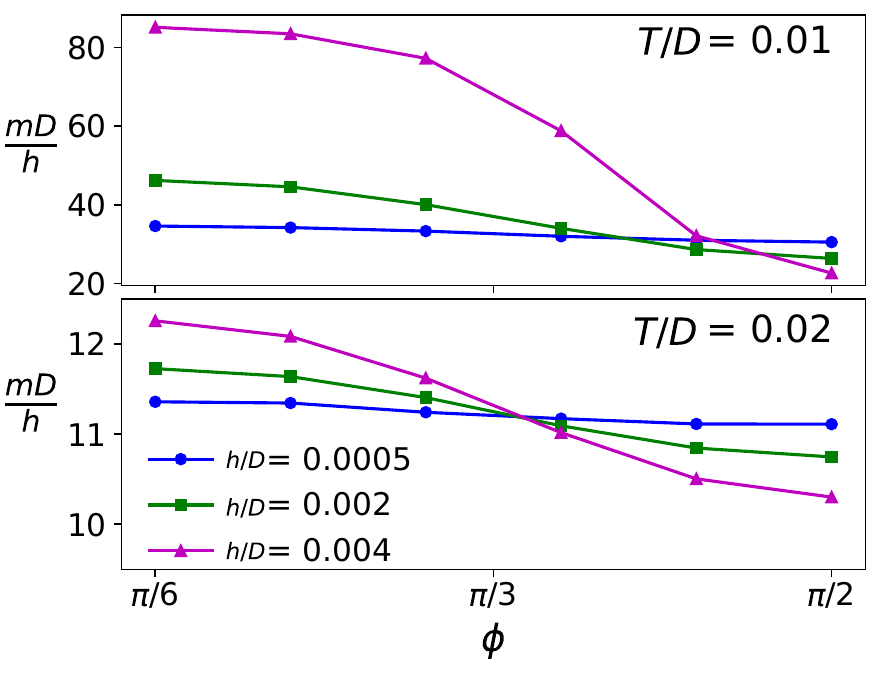}
\caption{
Rescaled magnetization \add{$ m D/h$} as a function of the direction of the external field $\phi$ in the system with \add{$U=300 D$} and $\delta \simeq 0.045$ at \add{$T = 0.01D$ (top) and $T=0.02D$ (bottom)}. Circles, squares, and triangles represent the results for \add{$h=0.0005D$, $0.002D$, and $0.004D$}, respectively.
}
\label{A_a}
\end{figure}

To discuss the magnetic instability in the paramagnetic state, we consider the magnetic susceptibility. To examine the magnetic response along the SU(3) principal axes, we introduce the two-dimensional external fields ${\bf h}=(h_3, h_8)$, and the corresponding Hamiltonian is given by
\begin{align}
 H_{\mathrm{ext}}=-{\bf h}\cdot{\bf S}, 
\end{align}
where ${\bf S}=(S_3, S_8)$. We note that, in general, the magnetization ${\bf m}$ is not parallel to the applied field ${\bf h}$. In fact, $m/h$ strongly depends on the direction of the magnetic field, specified by the angle $\phi(=\arg\,{\bf h})$, as shown in Fig.~\ref{A_a}. However, as the external field strength decreases, the directional dependence becomes negligible. In the limit $h\rightarrow 0$, the response becomes isotropic, allowing us to define the magnetic susceptibility as
\begin{equation}
\chi = \lim_{h\rightarrow 0} \frac{m}{h}.
\label{eq_chi}
\end{equation}
We confirm this isotropic behavior in a small external field ${\bf h}$ when the system is paramagnetic. In this study, we deduce the magnetic susceptibility, by examining the magnetization under a tiny field with $\mathrm{arg}\,{\bf m} =\pi/6$. In the following, we set the bandwidth $D$ as the energy unit.

\section{Numerical Results}\label{sec:r}

\begin{figure}[htb]
\includegraphics[width=0.9\linewidth]{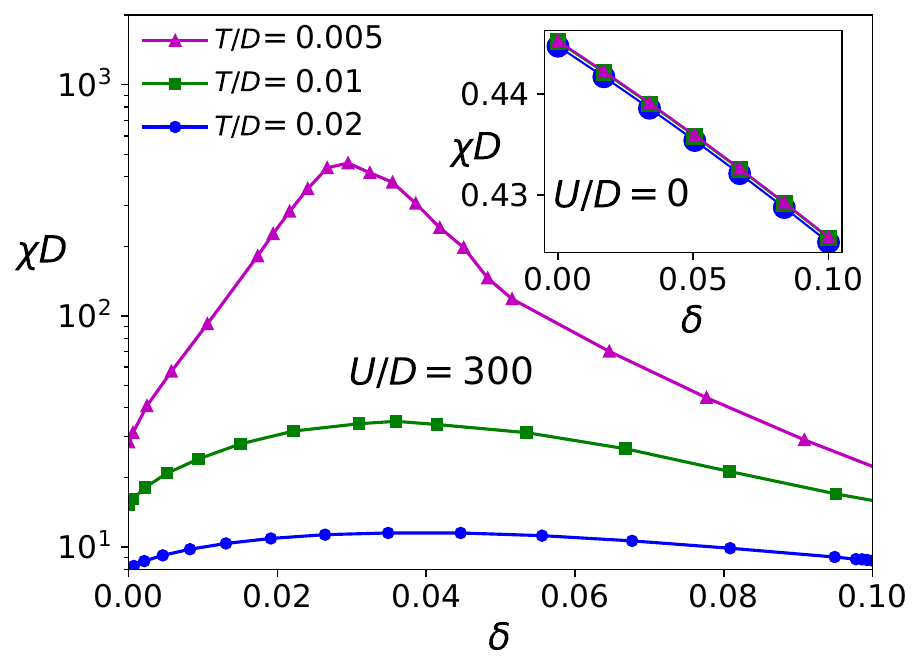}
\caption{
Magnetic susceptibility \add{$\chi D$} as a function of the hole doping $\delta$ in the system with \add{$U=300D$}. Triangles, squares, and circles represent the results for the temperatures \add{$T = 0.005D$, $0.01D$, and $0.02D$}, respectively. 
}
\label{A_b}
\end{figure}
We first examine the magnetic response in the SU(3) Fermi-Hubbard model. Figure~\ref{A_b} shows the doping dependence of the magnetic susceptibility of the system with the strong on-site interaction \add{$U=300D$}. 
\add{
Susceptibility is, in general, a key quantity for
identifying phase transitions.
Although it does not diverge at first-order transition points, nonmonotonic behavior of the susceptibility can serve as a precursor
to the magnetic phase transition.
} 
We find that the susceptibility exhibits a peak structure around $\delta\sim 0.03$ in contrast to the monotonic behavior observed in the noninteracting case, as shown in the inset of Fig.~\ref{A_b}. This nonmonotonic behavior originates from the strong on-site interaction. In addition, the peak structure becomes more pronounced as the temperature decreases. This indicates that magnetic fluctuations are enhanced in the SU(3) system although the susceptibility is isotropic. 
We confirm that this nonmonotonic behavior is absent in the SU(3) Fermi-Hubbard model on the Bethe lattice (see Appendix~\ref{bethe_} for details). A similar contrast has been reported in the SU(2) Fermi-Hubbard model~\cite{kamogawa_ferromagnetic_2019}, suggesting the emergence of Nagaoka-type ferromagnetism at low temperatures in the SU(3) case.

\begin{figure}[htb]
\includegraphics[width=0.99\linewidth]{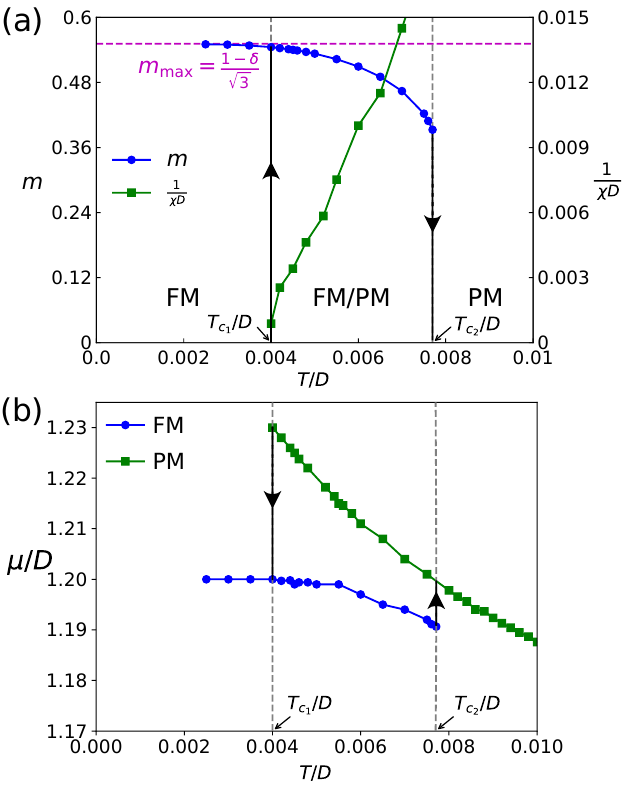}
\caption{
(a) Magnetization $m$ (left axis) and inverse susceptibility \add{$1/\chi D$} (right axis) as a function of temperature \add{$T /D$} in the system with \add{$U = 300 D$} and $\delta \simeq 0.045$. FM and PM indicate ferromagnetically ordered and paramagnetic states. (b) Solid circles (squares) represent the chemical potential for the FM (PM) solution. }
\label{B_a}
\end{figure}

To clarify this, we examine low-temperature properties in the SU(3) system with $\delta\simeq 0.045$. The results are shown in Fig.~\ref{B_a}(a). We find that the spontaneous symmetry breaking occurs accompanied by the emergence of magnetization at low temperatures. In this case, the direction of the magnetization is characterized as $\arg\,{\bf m} \sim \pi/6$, indicating that the polarized ferromagnetically ordered state with $n_1\sim 1-\delta$ and $n_2=n_3\sim 0$ is stabilized. We have confirmed that the direction of magnetization remains robust against changes in interaction strength and temperature, as long as the symmetry-broken state is realized. This state is essentially the same as the generalized Nagaoka ferromagnetically ordered state~\cite{katsura_2013}. As the temperature increases, the magnetization gradually decreases and suddenly drops to zero at \add{$T=T_{c_2}(\delta \simeq 0.045) \sim 0.0077D$}, suggesting the first-order phase transition to the paramagnetic state. In fact, as the temperature decreases, the paramagnetic state remains stable down to a certain temperature \add{$T=T_{c_1}(\delta \simeq 0.045) \sim 0.004D$}, below which the magnetization reemerges discontinuously. 
\add{
This hysteresis clearly indicates the presence of the first-order phase transition in the SU(3) system. In contrast, it has been reported within DMFT that the SU(2)-symmetric case exhibits a second-order transition between ferromagnetic and paramagnetic states~\cite{zitzler_magnetism_2002,kamogawa_ferromagnetic_2019,Koga_pro2021}. This difference suggests that the order of the transition depends on the symmetry of the system, even in strongly correlated regime, similarly to the prediction of Landau theory~\cite{cazalilla_2009}.
}
Also, Fig.~\ref{B_a}(b) shows the chemical potential for the system. We observe clear singularities accompanied by jumps at $T=T_{c_1}$ and $T_{c_2}$, which are consistent with the presence of the first-order phase transition.
\begin{figure}[htb]
\includegraphics[width=0.99\linewidth]{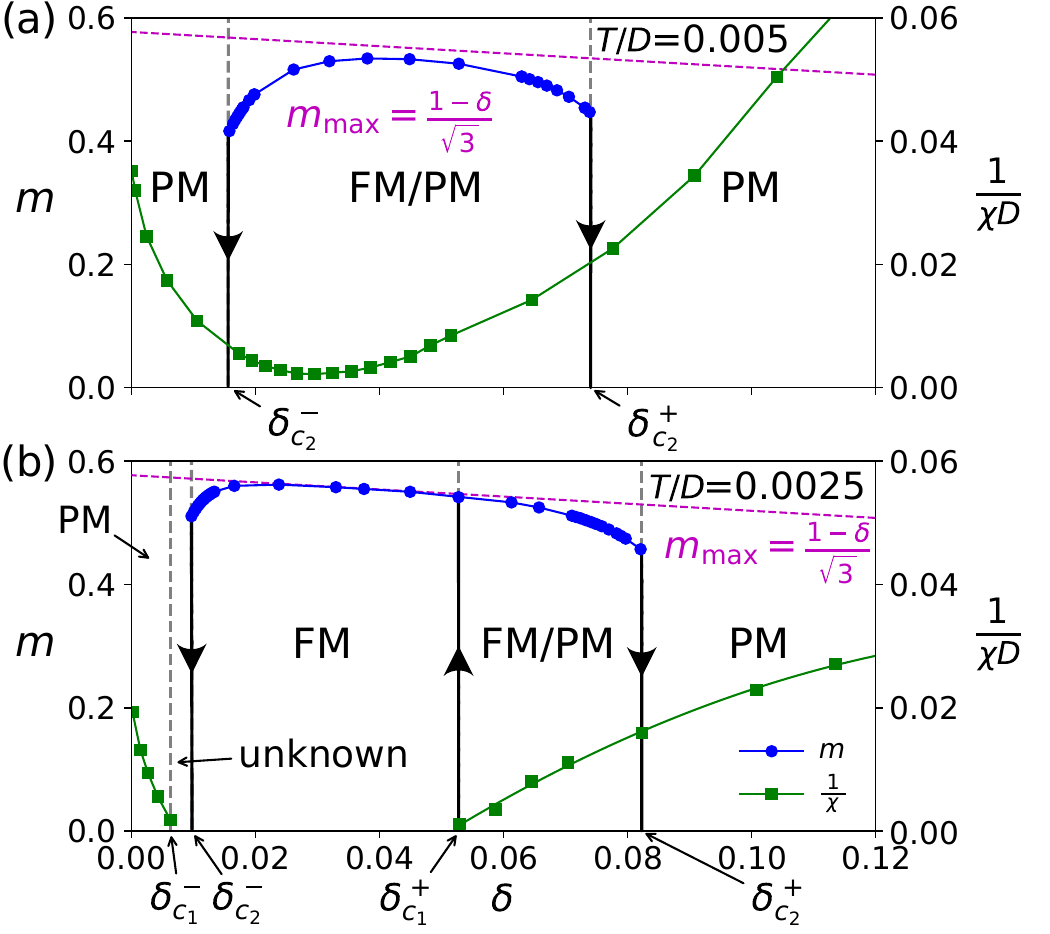}
\caption{
Magnetization $m$ (left axis) and inverse susceptibility \add{$1/\chi D$} (right axis) as a function of the hole doping $\delta$ at (a) \add{$T = 0.005D$ and (b) $T = 0.0025D$ for $U = 300D$}.
}
\label{C_a}
\end{figure}

Next, we clarify the stability of the ferromagnetically ordered state under hole doping.
Figure~\ref{C_a} shows the magnetization $m$ and the inverse of the magnetic susceptibility \add{$1/\chi D$ at $ U = 300 D$}. 
At a higher temperature ($ T = 0.005 $), the paramagnetic solution is always present, while the ferromagnetically ordered one appears only within the range $\delta_{c_2}^-<\delta<\delta_{c_2}^+$.
Therefore, when the doping rate varies within the paramagnetic regime, the ferromagnetically ordered state is not realized. Once the ferromagnetically ordered state is realized, the magnetization continuously decreases with changing the doping rate. Eventually, a first-order phase transition occurs accompanied by a discontinuous jump in the magnetization.

At a lower temperature \add{($T=0.0025D$)}, the genuine ferromagnetically ordered phase emerges when $\delta_{c_1}^-<\delta<\delta_{c_1}^+$, as shown in Fig.~\ref{C_a}(b). By increasing the doping rate, the first-order phase transition accompanied by hysteresis occurs. 
\add{In contrast, at a lower doping rate, approaching the transition point,
the physical quantities become highly sensitive to small changes of the chemical potential.
In this regime, the DMFT calculations are hard to converge due to enhanced particle-number fluctuations.}
\begin{figure}[htb]
\includegraphics[width=0.99\linewidth]{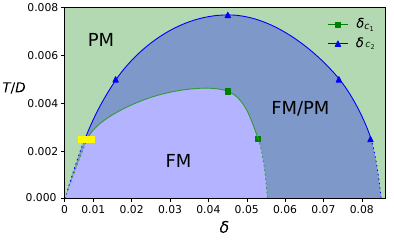}
\caption{
Phase diagram for the SU(3) Fermi-Hubbard model with \add{$U = 300 D$}. Solid squares (triangles) represent the phase transition points $\delta_{c_1}$ $(\delta_{c_2})$, where the paramagnetic (ferromagnetically ordered) solution disappears. The phase boundary is shown as a guide to the eye. 
A yellow rectangle indicates a region where the converged results were not obtained.
}
\label{C_b}
\end{figure}

Performing similar calculations, we obtain the phase diagram, as shown in Fig.~\ref{C_b}.
We find that the ferromagnetically ordered phase is realized at low temperatures and for low density of holes. This ferromagnetically ordered state is expected to be adiabatically connected to the Nagaoka limit ($T\rightarrow 0$ and $\delta\rightarrow 0$) although converged solutions are hard to obtain due to the enhanced particle-number fluctuations.

\begin{figure}[htb]
\includegraphics[width=0.99\linewidth]{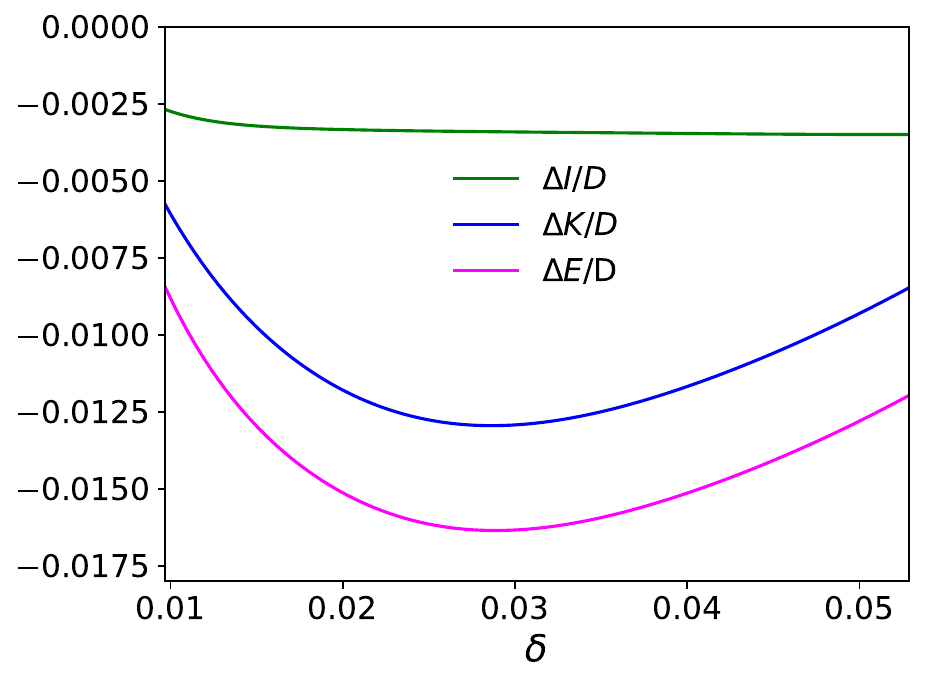}
\caption{
Condensation energy \add{$\Delta E /D$} and its contribution \add{$\Delta K / D$} and \add{$\Delta I / D$} as a function of the hole doping in the ferromagnetically ordered state when \add{$U=300D$} and \add{$T = 0.0025D$}.
}
\label{ene}
\end{figure}

We now discuss the driving mechanism that stabilizes the ferromagnetically ordered state in the SU(3) Fermi-Hubbard model. To this end, we calculate the kinetic energy and interaction energy. The kinetic energy per site is given by $K=-2\pi T\sum_{n, \sigma} G_{\mathrm{imp}}^\sigma(i\omega_n)\Delta^\sigma(i\omega_n)$, the interaction energy per site is given by $I=U/2\sum_{\sigma\neq\sigma'}\expval{n_{\sigma}n_{\sigma'}}$, and the total internal energy per site is thus expressed as $E=K+I$. For comparison, we also calculate these energy contributions in the paramagnetic state, where the condition $n_1=n_2=n_3$ is imposed. Figure~\ref{ene} shows the condensation energy $\Delta E=E_{\mathrm{FM}}-E_{\mathrm{PM}}$, and their contributions $\Delta K=K_{\mathrm{FM}}-K_{\mathrm{PM}}$ and $\Delta I=I_{\mathrm{FM}}-I_{\mathrm{PM}}$, in the system with \add{$U=300 D$ and $T=0.0025 D$}. As $n_1 \sim 1-\delta$ and $n_2=n_3\sim 0$ are realized in the ferromagnetically ordered state, the relation $\Delta I\sim -I_{\mathrm{PM}}$ is satisfied and it shows only a weak dependence on the doping rate. We then find that the ferromagnetically ordered state is energetically favored over the paramagnetic state owing to the dominant contribution of the kinetic energy. Additionally, in the small $\delta$ case, the energy gain is approximately proportional to the doping rate, which is consistent with the fact that the Nagaoka ferromagnetism is stabilized by the kinetic energy of the doped holes. As $\delta$ increases, $\Delta E$ reaches a minimum around $\delta\sim 0.03$, which corresponds to the maximum magnetization in this parameter region.

\begin{figure}[htb]
\includegraphics[width=0.9\linewidth]{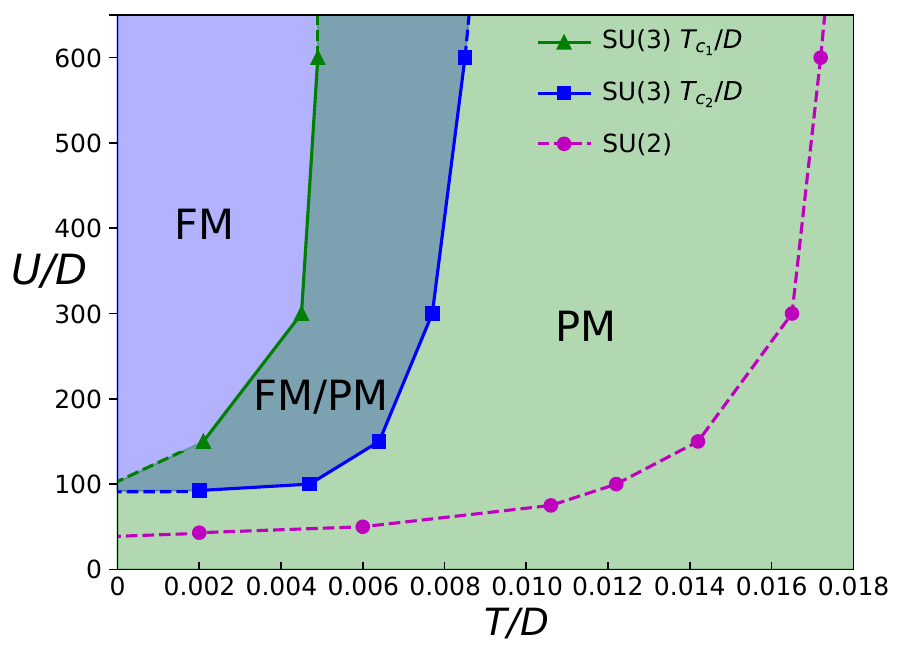}
\caption{
$U$-$T$ phase diagrams for the SU(2) and SU(3) Fermi-Hubbard model with $\delta\simeq 0.045$. The green and blue curves correspond to the temperatures $T=T_{c_1}$ and $T=T_{c_2}$, where the paramagnetic and ferromagnetically ordered solutions disappear, respectively.
}
  \label{E_a}
\end{figure}

Finally, we compare our results with those of the SU(2) Fermi-Hubbard model. Figure~\ref{E_a} shows the $U$-$T$ phase diagrams for the SU(2) and SU(3) Fermi-Hubbard models with $ \delta \simeq 0.045 $. In the SU(2) case, the magnetic phase transition is of second order~\cite{obermeier_ferromagnetism_1997,kamogawa_ferromagnetic_2019,Koga_pro2021}, and thereby critical behavior appears at the phase boundary. In contrast, the SU(3) system exhibits hysteresis behavior characteristic of a first-order phase transition. Notably, the SU(3) ferromagnetically ordered phase emerges at lower temperatures and requires the stronger interaction strength, in contrast to the SU(2) case. This is qualitatively consistent with the results for the single hole doped Fermi-Hubbard model on the square lattice~\cite{botzung_numerical_2024}.

\add{
In our analysis, we have restricted our discussions to the ferromagnetic instability,
and have not considered other ordered states such as antiferromagnetically ordered state~\cite{jarrell1992hubbard,snoek2008antiferromagnetic,rampon2025magnetic}, canted state~\cite{snoek2011canted,sotnikov2013magnetic,Ueda2012Electronic,Gradi2018Correlation}, excitonic state~\cite{Kune2014Excitonic,Geffroy2019Collective}, etc.
This is because even the stability of a simple ordered state characterized 
only by diagonal components has not been investigated in the SU(3) case.
It is interesting to clarify how such ordered states compete or coexist
with the ferromagnetically ordered state,
which will be discussed elsewhere.}

\section{Conclusions}\label{sec:conclusion}
We have investigated the SU(3) Fermi-Hubbard model on the hypercubic lattice, using DMFT combined with the CT-QMC method. By analyzing the magnetization and magnetic susceptibility -- particularly their dependence on temperature and hole doping -- we have demonstrated that a ferromagnetically ordered state is realized in the strong-coupling and low-doping regime. Moreover, we have identified hysteresis behavior as temperature and/or doping are varied, indicating the presence of a first-order phase transition. Furthermore, by examining the kinetic and interaction energies, we have confirmed that the kinetic energy gain plays a dominant role in stabilizing this ordered state, which is consistent with its adiabatic connection to the Nagaoka ferromagnetism. Compared to the SU(2) case, the ferromagnetically ordered phase emerges at lower temperatures and requires stronger interaction strengths in the SU(3) system.

Multicomponent systems have been realized in ultracold atoms, and thereby it is worth exploring the ferromagnetism in systems with larger internal degrees of freedom, such as $N=3$, $N=6$ and $N=10$, which have been realized using $^6$Li~\cite{ottenstein_collisional_2008,Huckans2009three}, $^{87}$Sr~ \cite{desalvo_degenerate_2010,zhang2014spectroscopic}, and $^{173}$Yb
\add{~\cite{Fukuhara2007degenerate,taie2010realization,pasqualetti2024EOS}}, 
respectively. In addition, studying the spatial fluctuations that arise in finite-temperature ferromagnetically ordered states remains an intriguing subject. As the ferromagnetic phase is stabilized by the formation of a Nagaoka polaron in the ground state~\cite{samajdar_polaronic_2024}, it is interesting to determine whether a similar quasiparticle picture persists at finite temperatures. We expect that both spin inhomogeneity and the distribution of hole positions may play an important role in the stabilization mechanism.

\begin{acknowledgments}
This work was supported by JSPS KAKENHI Grants No.\ JP22K03525, JP25H01521, JP25H01398 (A.K.), and No.\ JP25K17327 (K.Y.). K.Y. was also supported by Murata Science and Education Foundation, Hirose Foundation, the Precise Measurement Technology Promotion Foundation, and the Fujikura Foundation. Parts of the numerical calculations were performed in the supercomputing systems in ISSP, the University of Tokyo. Parts of the simulations have been performed using the ALPS libraries~\cite{bauer_alps_2011}.
\end{acknowledgments}

\appendix
\section{Nonuniform sampling scheme in CT-QMC}\label{non_uniform_sampling}

\begin{figure}[htb]
  \includegraphics[width=0.9\linewidth]{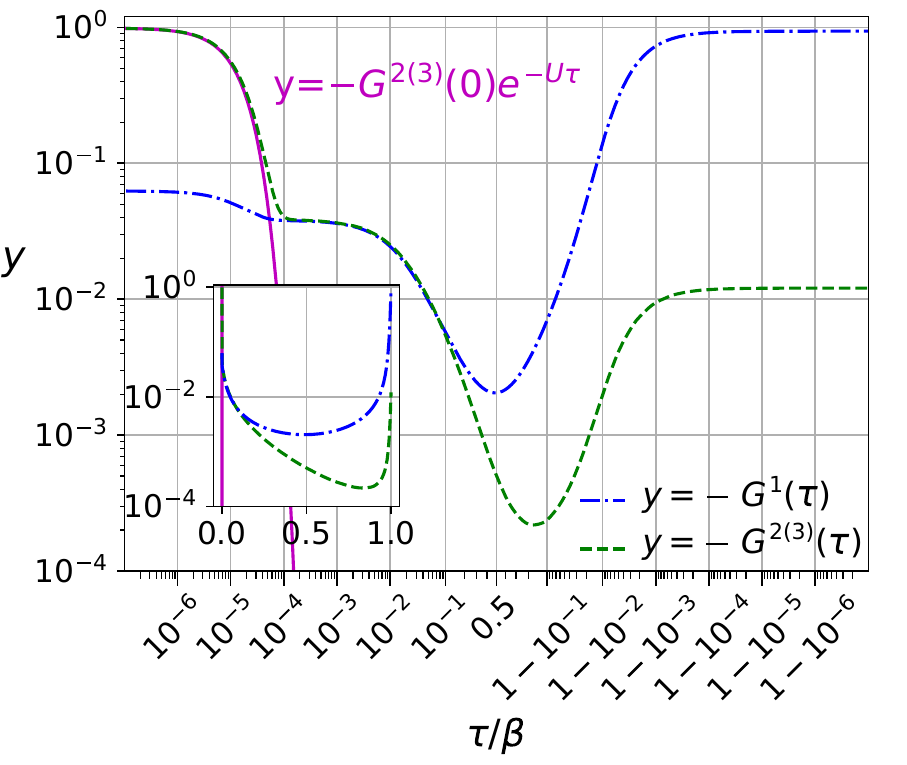}
  \caption{Imaginary-time Green's function $G^{\sigma}(\tau)$ as a function of $\tau / \beta$ for \add{$U = 300 D$, $T = 0.005 D$}, and $\delta \simeq 0.9619$ (\add{$\mu  = 1.25D$}) in the ferromagnetically ordered phase. The blue and green curves correspond to the internal states with large and small particle number density, respectively. The solid cyan line represents the exponential decay as $y=-G^{2(3)}(0) \exp(-U \tau)$. The horizontal axis is scaled logarithmically, while the inset presents the same data on a linear scale for comparison.}
  \label{E_b}
\end{figure}

We explain the details of the nonuniform modification used in performing the sampling of Green's functions in the segment-based CT-QMC method, where the Green's function $G^{\sigma}(\tau)$ is obtained as
\begin{equation}
  G^{\sigma}(\tau) 
  = \frac{1}{\beta} 
    \left\langle 
      \sum_{i,j}
      O_{ji}^{\sigma}\,\mathrm{sgn}\left(\tau_{e,i}^{\sigma} - \tau_{s,j}^{\sigma}\right)\,
      \delta\!\left(\tau - (\tau_{e,i}^{\sigma} - \tau_{s,j}^{\sigma})\right)
    \right\rangle_{\mathrm{MC}}.
  \label{eq_sampling}
\end{equation}
Here, $\tau_{s,i}^\sigma$ and $\tau_{e,j}^\sigma$ represent the start and end times of each segment for spin $\sigma$, respectively~\cite{werner_continuous_2006,gull_continuous_2011}, $O_{i,j}^\sigma$ is defined by using the hybridization function $\Delta^\sigma(\tau)$ as
\begin{equation}
  O_{i,j}^{\sigma} 
  = \Delta^{\sigma}\left(\tau_{s,i}^{\sigma} - \tau_{e,j}^{\sigma}\right),
\end{equation}
and $\langle\cdot\rangle_{\mathrm{MC}}$ denotes the Monte Carlo average. In general, when we have $M$ sampling points in the interval $(0,\beta)$, the delta function is defined as 
\begin{equation}
  \delta(\tau) = 
  \begin{cases}
    \displaystyle \frac{M}{\beta}, & -\frac{\beta}{2M} \le \tau \le \frac{\beta}{2M}, \\
    0, & \text{otherwise.}
  \end{cases}
\end{equation}
However, we find that equally spaced sampling scheme encounters numerical difficulties in the strongly correlated regime, where the Green's function exhibits an exponential decay near $ \tau \simeq 0, \beta $ and takes extremely small positive values around $\tau \simeq \beta/2$ as shown in Fig.~\ref{E_b}. To accurately capture the exponential decay near $\tau=0$, we have to reduce the discretization errors by increasing the number of sampling points $M$. On the other hand, to obtain sufficiently precise values around $\tau\sim \beta/2$, where the Green's function is very small, it is crucial to increase the sampling counts per point to reduce the standard error. A simple improvement would be to enlarge the sampling interval so as to collect more samples at each point, which in turn leads to a reduction in $M$. Consequently, equally spaced sampling scheme becomes inefficient for the calculation.

Instead, we use a nonuniform sampling method that overcomes this issue by allowing fine samplings near the boundaries and coarse samplings around the center, thereby achieving an efficient global representation of the Green's function. In this work, we choose the following sampling points
\begin{equation}
  \tau_i 
  = \beta \,\sin^2\!\left(\tfrac{\pi}{2}\,\tfrac{i}{M-1}\right).
  \label{tau_i}
\end{equation}
This choice ensures that the sampling remains nearly uniform in the central region, whereas the sampling interval becomes finer in the vicinity of the boundaries. We then introduce a set of points $\{\xi_i\}$ that satisfies
\begin{equation}
  \tau_i = \frac{\xi_i + \xi_{i+1}}{2},
  \label{sep}
\end{equation}
which gives the delta function as
\begin{equation}
  \delta(\tau - \tau_i) 
  = \begin{cases}
      \frac{1}{\xi_{i+1} - \xi_i}, & \xi_i \le \tau \le \xi_{i+1}, \\[5pt]
      0, & \text{otherwise}.
    \end{cases}
\end{equation}
Furthermore, $\xi_i$ is analytically calculated as
\begin{equation}
  \xi_i 
  = \frac{\beta}{2}\,\left(
       1 
       - \frac{\cos\!\left[\left(i - \frac{1}{2}\right)\,\tfrac{\pi}{M - 1}\right]}
               {\cos\!\left(\tfrac{1}{2}\,\tfrac{\pi}{M - 1}\right)}
     \right).
\end{equation}
This provides an explicit expression for the sampling boundaries of the delta function. With the use of this protocol, we find that the sampling intervals for $G^{\sigma}(0)$ and $G^{\sigma}(\beta)$ are not included, and these values cannot be directly obtained from Eq.~(\ref{eq_sampling}). Instead, they are determined from the particle number as 
\begin{equation}
    1-G^{\sigma}(0) = G^{\sigma}(\beta) = -\expval{n_{\sigma}},
\end{equation}
which allows the entire Green's function to be reconstructed. In our numerical calculations, we set $M = 10{,}000$, which leads to $\tau_1/\beta = 2.47 \times 10^{-8} \sim 1/M^2$ and $(\tau_{M/2} - \tau_{M/2-1})/\beta = 1.57 \times 10^{-4} \sim 1/M$. 

In addition, the nonuniform sampling protocol defined by Eq.~\eqref{tau_i} provides an advantage in evaluating the hybridization function. In the self-consistent DMFT calculation, the hybridization function tends to acquire a similar functional structure as the Green's function~\cite{georges_dynamical_1996}, and therefore, it is crucial to represent it by using the nonuniform sampling scheme to improve the accuracy. During the CT-QMC simulation, the value of the hybridization function is evaluated repeatedly. This requires to determine the corresponding array index from a given imaginary time $\tau$, which in general necessitates a binary search. However, from Eq.~\eqref{tau_i}, we can analytically obtain
\begin{equation}
    i = \left\lfloor \frac{2(M-1)}{\pi} \sin^{-1}\left(\sqrt{\frac{\tau}{\beta}}\right) \right\rfloor,
\end{equation}
where $\left\lfloor \cdot \right\rfloor$ denotes the floor function. This inverse mapping enables direct index computation without searching for it, leading to almost the same computational efficiency as that in the uniform case.

Finally, we note that our nonuniform sampling protocol and the general procedure of constructing sparse sampling points from compact orthogonal basis functions are closely related~\cite{li_sparse_2020}. While the Fourier transformation via the intermediate-representation (IR) method is efficient, using uniformly spaced grids in either the imaginary-time or Matsubara-frequency domain results in ill-conditioned transforms due to the near-linear dependence among the basis vectors. Reference~\cite{li_sparse_2020} proposes to improve the conditioning by selecting sampling points as the roots of the basis functions. Interestingly, although our sampling points were designed to achieve a globally accurate representation of the Green's function, they are found to lie close to the roots of the IR basis functions, which exhibit strong oscillations near the boundaries and weak oscillations near the center. Therefore, the proposed sampling points enable a stable and accurate transformation to the IR basis.

\section{Magnetic susceptibility for the Bethe lattice}\label{bethe_}
Here, we briefly discuss the absence of ferromagnetism on the Bethe lattice, which has an infinite coordination number and DMFT calculations also become exact. The noninteracting density of states (DOS) is given by
\begin{equation}
    \rho(\epsilon) = \frac{2}{\pi D} \sqrt{1 - \left(\frac{\epsilon}{D}\right)^2},
\end{equation}
which leads to the simple Dyson equation of the effective impurity model,
\add{
\begin{equation}
    \add{\mathcal{G}^{\sigma}(i\omega_n)^{-1}} = i\omega_n + \mu - \frac{D^2}{4}G^{\sigma}_{\mathrm{loc}}(i\omega_n).
\end{equation}
By combining this expression with Eq.~\eqref{eq_imp}, we obtain the simple self-consistent equation $\Delta^{\sigma}(\tau) = \frac{D^2}{4}G_{\mathrm{loc}}^{\sigma}(\tau)$, which allows the DMFT calculation to be carried out entirely in imaginary time.}
We remark that Nagaoka's theorem does not apply to the Bethe lattice, as it lacks the closed-loop structures required for the theorem. 

\begin{figure}[b]
  \includegraphics[width=0.9\linewidth]{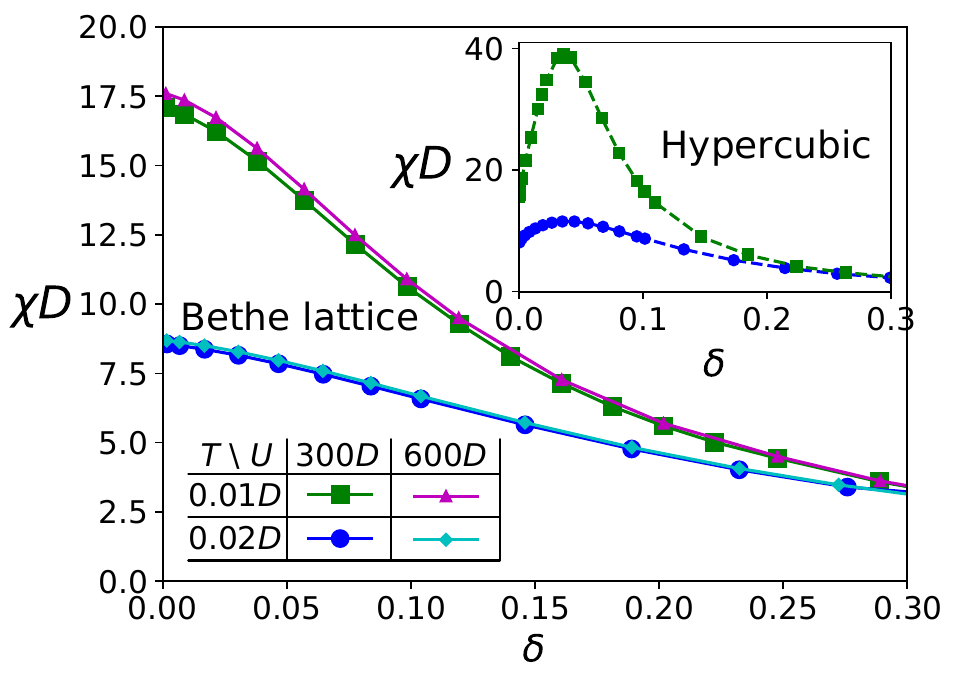}
  \caption{Magnetic susceptibility \add{$\chi D$} as a function of the hole doping $\delta$ for the Bethe lattice (inset: for the hypercubic lattice in infinite dimensions). The plots are for \add{$U = 300D$ at $T = 0.01D$ (green) and $T = 0.02D$ (blue) on both hypercubic and Bethe lattices, and $U = 600D$ at $T = 0.01D$ (magenta) and $T = 0.02D$} (cyan) on the Bethe lattice. The absence of a peak structure indicates that the ferromagnetically ordered state is unstable on the Bethe lattice.}
  \label{F_a}
\end{figure}

Figure \ref{F_a} shows the magnetic susceptibility \add{$\chi D$}
as a function of the hole doping $ \delta $ in the strongly correlated regime at low temperatures, where the SU(3) ferromagnetism emerges in the case of the infinite-$d$ hypercubic lattice (see the peak structure in the inset). Clearly, we find that the magnetic susceptibility exhibits no peak on the Bethe lattice. As discussed in the main text, the peak structure in the susceptibility signals the onset of the ferromagnetic order, and thereby the absence of a peak structure indicates that the ferromagnetism does not emerge on the Bethe lattice. This result highlights that the lattice geometry is crucial for stabilizing itinerant ferromagnetism in the SU(3) Fermi-Hubbard model.

We note that the absence of the ferromagnetism on the Bethe lattice has been also reported in the SU(2) Fermi-Hubbard model on the basis of the DMFT calculations~\cite{kamogawa_ferromagnetic_2019}, where no spontaneous magnetization is observed. In Ref.~\cite{kamogawa_ferromagnetic_2019}, it is discussed that the absence of high-frequency components in the noninteracting DOS may be a possible origin of the absence of the ferromagnetism. As the SU(3) ferromagnetism in the current study is of kinetic origin as in the case of the SU(2) ferromagnetism, such an argument may hold for the SU(3) Fermi-Hubbard model, but clarifying the precise structure is left for future study.

%

\end{document}